\documentclass[aps,prl,reprint,showpacs,amsmath,amssymb,superscriptaddress]{revtex4-1}
\usepackage{graphicx}

\begin{document}

\title{Duality of Weak and Strong Scatterer in Luttinger liquid Coupled to Massless Bosons}

\author{Igor V.\ Yurkevich}
\affiliation{Nonlinearity and Complexity Research Group, Aston University, Birmingham B4 7ET, United Kingdom}
\affiliation{School of Physics and Astronomy, University of Birmingham,
Birmingham B15 2TT, United Kingdom}
\affiliation{The Abdus Salam International Centre for Theoretical Physics, P.O. Box 586, 34100 Trieste, Italy}
\affiliation{Institut f\"{u}r Nanotechnologie, Karlsruhe Institute of Technology, D-76021 Karlsruhe, Germany}

\author{Alexey \ Galda}
\affiliation{School of Physics and Astronomy, University of Birmingham,
Birmingham B15 2TT, United Kingdom}
\affiliation{Materials Science Division, Argonne National Laboratory, Argonne, Illinois 60439, USA}
\author{Oleg M.\ Yevtushenko}
\affiliation{Arnold Sommerfeld Center and Center for Nano-Science, Ludwig Maximilians University, Munich, DE-80333, Germany}
\affiliation{The Abdus Salam International Centre for Theoretical Physics, P.O. Box 586, 34100 Trieste, Italy}
\author{Igor V.\ Lerner}
\affiliation{School of Physics and Astronomy, University of Birmingham,
Birmingham B15 2TT, United Kingdom}
\affiliation{The Abdus Salam International Centre for Theoretical Physics, P.O. Box 586, 34100 Trieste, Italy}
\def\sgn{\operatorname{sgn}}

\begin{abstract}

We study electronic transport in a Luttinger liquid (LL) with an embedded impurity, which is either a weak scatterer (WS) or a weak link (WL), when interacting electrons are coupled to one-dimensional massless bosons (e.g., acoustic phonons). We find that the duality relation, $\Delta _{{\text{WS}}}\Delta _{\text{WL}}=1 $, between scaling dimensions of the electron backscattering in the WS and WL limits,  established for the standard LL,  holds  in the presence of the additional coupling for an arbitrary \emph{fixed} strength of boson scattering from the impurity. This means that at low temperatures such a system remains either an ideal insulator or an ideal metal, regardless of the scattering strength. On the other hand, when fermion and boson scattering from the impurity are correlated, the system has a  rich phase diagram that includes a metal-insulator transition at some intermediate values of the scattering.
\end{abstract}
\pacs{71.10.Pm, %Fermions in reduced dimensions (anyons, composite fermions, Luttinger liquid, etc.)
 73.63.Nm %Quantum wires
}

\maketitle

Low-temperature physics of one-dimensional electron systems, like quantum wires or nanotubes, is governed by electron-electron (e-e) interactions. Electrons in such systems form a Luttinger liquid (LL) \cite{Tom:50,*Lutt:63,*HALDANE:81} characterized by a power-law decay of various correlation functions (see Refs.~\cite{CazalillaRMP:11,*1Dreview:10,Giamarchi,GogNersTsv,vDSh:98} for reviews), which has been experimentally revealed via conductance measurements and a scanning tunneling microscopy both in carbon nanotubes \cite{Bockrath:99,*Yao:99,*Ishii:03,*Lee:04} and quantum nanowires \cite{Auslaender:02,*Slot:04,*Levy:06,*Kim:06}. In particular, inserting a single impurity or a weak link (e.g., a tunnel barrier) into a LL leads at low temperatures $T$ to the power-law suppression of the conductance through the system and of a local density of states at the impurity site
\cite{KaneFis:92a,*KF:92b,MatYueGlaz:93,FurusakiNagaosa:93b,Furusaki:97,%
EggertAffleck:92,*EggertAffleck:95,*FabrizioGogolin:95} with the latter fading away with the distance \cite{Eggert:00,GYL:04}.

The low-$T$ suppression of conductance is caused by a power-law enhancement of a backscattering amplitude $\lambda $ from the impurity at low energies $\varepsilon   $ \cite{KaneFis:92a}, $\lambda ({\varepsilon   })\sim \lambda\, \varepsilon   ^{\Delta _\mathrm{WS}-1 }  $. Here the weak scattering scaling dimension    $\Delta _\mathrm{WS} =K$ where the Luttinger parameter $K$ is smaller than $1$  in the LL with an $e$-$e$ repulsion. It was argued \cite{KaneFis:92a} that the limit of strong scattering is equivalent to a weak link  with a small  tunneling amplitude $ t_\mathrm{WL}$  between two semi-infinite wires, which is suppressed in the low-energy limit   as $t_\mathrm{WL}({\varepsilon   })  \sim t_\mathrm{WL}\, \varepsilon   ^{\Delta _\mathrm{WL}-1 } $. The scaling dimensions  $\Delta_{\mathrm{WS}}$ and $\Delta_{\mathrm{WL}}$ obey the  duality relation,
\begin{align}\label{Dual}
   \Delta_{\mathrm{WS}} \, \Delta_{\mathrm{WL}}=1\,.
\end{align}
Thus, when weak scattering is a relevant perturbation, weak tunneling is an irrelevant one. This means that zero conductance (no tunneling) corresponds to a stable fixed point for renormalization group (RG) flows, while zero scattering (i.e.\ a perfect conductance of $e^2/h$ per channel \cite{MasStone:95,*Ponomarenko:95,*SafiSch:95}) to an unstable fixed point.  The relation ({\ref{Dual}}) holds also when $K>1$, i.e.\ in the LL of fermions with attraction or bosons with repulsion, but the direction of the RG flows reverses there \cite{KaneFis:92a}. Therefore, in a low-$T$ limit the LL is either an insulator or an ideal conductor, regardless of the bare value of $ \lambda $ or $  t_\mathrm{WL}  $. This RG prediction has been confirmed for an arbitrary impurity strength by a perturbative calculation for weakly interacting fermions \cite{MatYueGlaz:93,Aristov:09}, as well as by an exact calculation at $K={1}/{2} $ \cite{vDSh:98,Furusaki:97}. Similar approaches also work for more complicated defect structure (a resonant or side-attached impurity, a double-barrier structure, etc.) \cite{KaneFis:92b,FurMatv:02,*NazGlaz:03,*PolGorn:03,LYY:08,*GB:10}.

The duality relation ({\ref{Dual}}), which underpins the character of RG flows, is robust within the standard Tomonaga-Luttinger (TL) model of interacting electrons with a linearized spectrum. Originally \cite{KaneFis:92a} it was shown to follow from the duality of fields whose correlation functions yield the scaling dimensions  $\Delta_{\mathrm{WS}}$ and $\Delta_{\mathrm{WL}}$.  It was stated later \cite{FendleyLudwigSaleur:95,*FendleyLudwigSaleur:95a,*FendleySaleur:98} that the duality holds due to integrability of the TL model with a weak or strong scatterer. A natural question to ask is whether the duality still holds for realistic quantum wires or nanotubes, where additional interactions might break down the integrability?

In the present Letter  we address this question by considering the LL coupled to massless bosons thus modeling an unavoidable interaction of electrons with acoustic phonons. In the low-energy limit, an effective (i.e.\ mediated by phonons) $e$-$e$ interaction is retarded and thus cannot be reduced to a renormalization of parameters of the TL model. Then the scaling dimensions $\Delta_{\mathrm{ws,wl}}$ depend  on a number of additional parameters: a strength of the electron-phonon ($e$-ph) coupling, $g_\mathrm{ph}$, the ratio of the electron excitations (i.e.\ plasmon) velocity to that of sound, $\beta=v/c$,  and  finally  on a  backscattering amplitude $ r $ of phonons from the impurity (ranging from $0$ to $-1$). Without referring to the integrability (as there is no evidence that it survives coupling to phonons), the existence of any meaningful relation between $\Delta_{\mathrm{WS}}$ and $\Delta_{\mathrm{WL}}$, not speaking of the duality,  seems \emph{a priori} to be rather unlikely.

Nevertheless, a straightforward calculation presented here shows that the  the duality (\ref{Dual}) remains valid for an arbitrary set of the parameters listed above, albeit it is considerably more complicated than the change $K\to1/K$ in the standard TL model:
\begin{align}\label{Delta}
\Delta^{-1}_{\mathrm{WL}}=\Delta_{\mathrm{WS}}=K\,\frac{(1+{r})(1+\beta \kappa)-{r} W}{(1+{r})W\kappa-{r} (\kappa+\beta)}
\end{align}
with $\kappa\!\equiv\!\sqrt{1-\alpha}$,  $W\!\equiv\!\sqrt{1+2\beta \kappa + \beta^2}$, and  $\alpha\!\equiv\! g_\mathrm{ph}^2K/\pi v$.

Equation (\ref{Delta}) is our main result, obtained analytically by a ``brute force".  We are not currently aware of any symmetry responsible for this and cannot state whether the duality extends beyond the relation (\ref{Dual}) for scaling dimensions.

\begin{figure}  {\includegraphics[width=0.85\columnwidth,clip]{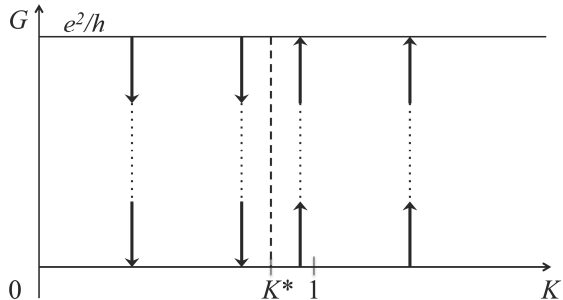}}\caption{\label{Fig1}RG flows  for fixed phonon parameters, assuming that electron and phonon scattering from the impurity are uncorrelated. A transition from insulator ({$G\to0$}) to metal ($G\to e^2/h$) is shifted from  $K =1$ to a new threshold  value $K^*$ which depends on phonon parameters.}
   \end{figure}

Speaking about experimental signatures of the duality in the presence of the $e$-ph coupling, it is important to stress that there can be two principally different situations, depending on whether the scattering properties of electrons and phonons from a single defect are  correlated or not.  The latter is realized, for example,  by locally depleting electron density at the impurity by  a charged plunger.  In this case, the phonon scattering is not changed during a crossover between the WS and WL limits. The duality ({\ref{Dual}}) means that the direction of RG flows is the same in both limits, see Fig.\ref{Fig1}. The only difference from the original picture \cite{KaneFis:92a} is that the flow direction changes at some point $K^*<1$ since the el-el repulsion is weakened by the phonon-mediated attraction.

\begin{figure}[t]  {\includegraphics[width=.85\columnwidth,clip]{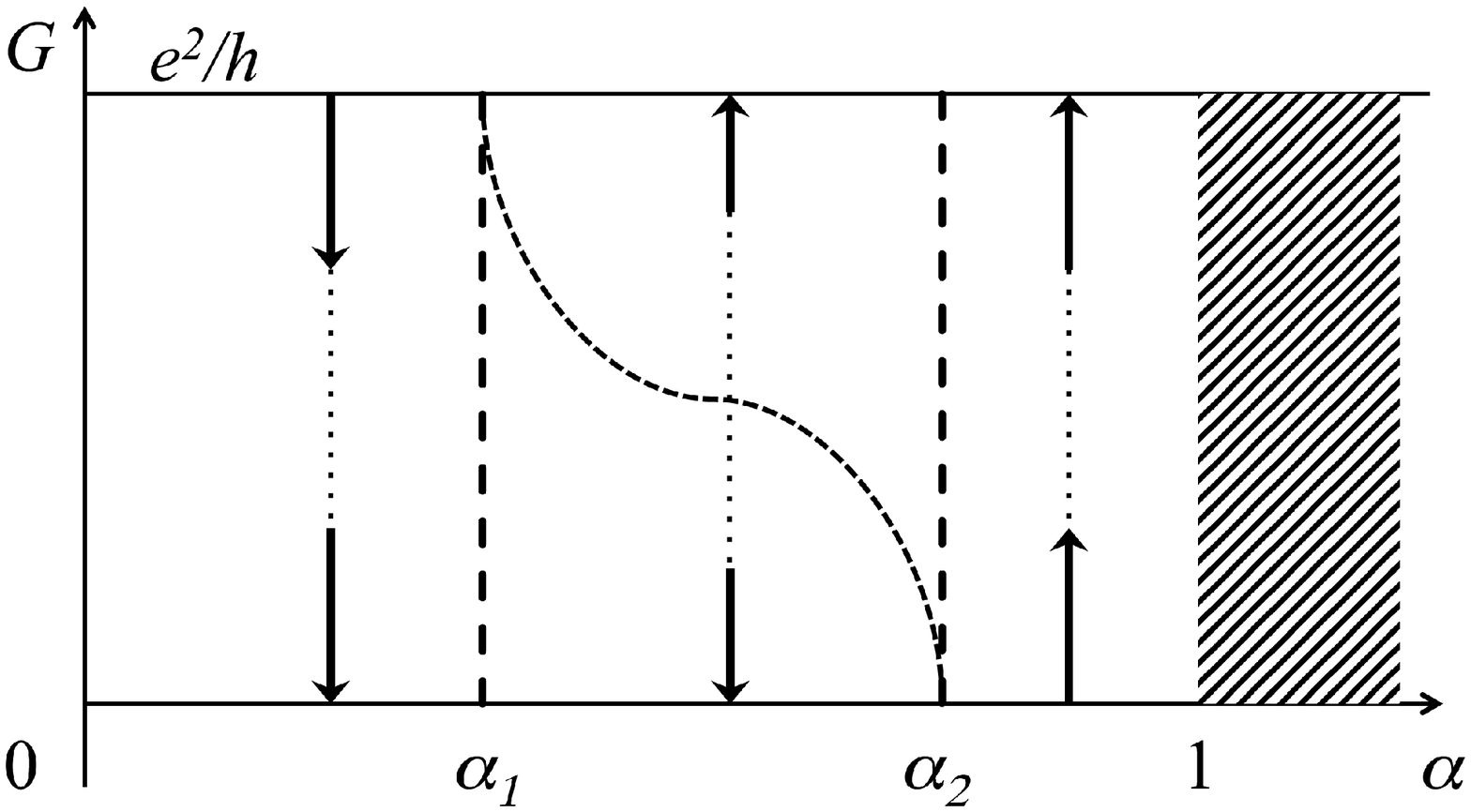}} \\ \centerline{\textbf{(a)}} {\includegraphics[width=.9\columnwidth,clip]{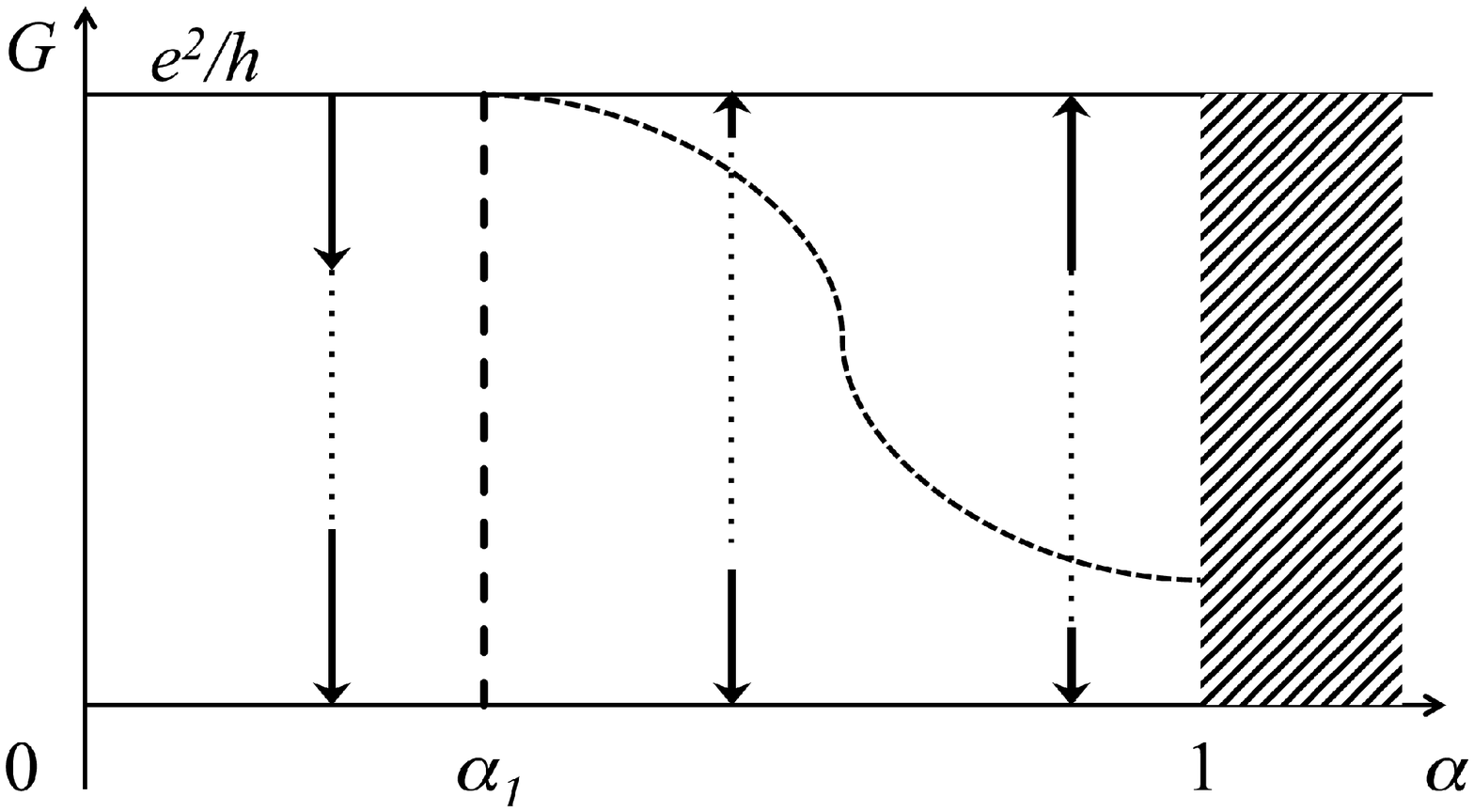}}\\ \centerline{\textbf{(b)}}\caption{\label{Fig2}RG flows for the correlated electron and phonon scattering from the impurity at different values of the dimensionless coupling $\alpha $. We assume that both electrons and phonons are almost fully transmitted through the impurity in the WS limit and almost fully backscattered in the WL limit: then $\Delta _\mathrm{WS}\!=\!1 $ at some $\alpha\!=\!\alpha_1$ while $\Delta _\mathrm{WL}\!=\!1$ at  $\alpha \!=\!\alpha_2\!\ne\!\alpha_1$.  For   $\alpha_1<\alpha<\min\{{\alpha_2,1}\} $ the insulator ({$G=0$}) and metal ({$G=e^2/h$}) fixed points are both stable. Thus a line of unstable fixed points corresponding to a metal-insulator transition (MIT) at each $\alpha$ in this interval should exist at finite $G<e^2/h$  at some intermediate value of the bare backscattering. Depending on a value of $\beta$, such a line might  (a) end at $G=0$ dividing the phase diagram in the regions of insulator, $\alpha\le\alpha_1$, metal, $\alpha_2\le\alpha<1$, or MIT, $\alpha_1<\alpha<\alpha_2$; or ({b}) end at the Wentzel-Bardeen instability line, $\alpha=1$, in which case the purely metallic region is absent.}
\end{figure}

On the other hand, both scattering strengths can be changed in parallel,
e.g., by bending a suspended nano\-tube or by inducing local structural change
with a tip of an atomic force microscope.  The duality relation (\ref{Dual}) does
not apply to this case since $\Delta _\mathrm{WS} $ and $\Delta _\mathrm{WL} $ must be taken at different values of a phonon backscattering amplitude from the impurity.   Thus there exists a certain range of parameters characterizing phonon propagation where both weak  scattering and tunneling through a weak link become irrelevant (both $\Delta $ are larger than $1$). As the RG flows have opposite directions in this region, there should exist a line of fixed points separating  the flows to the insulating fixed points ($G\to0$) from those to the metallic ones ({$G\to e^2/h$}),
see Fig.\ref{Fig2}. This indicates the existence of a metal-insulator
transition controlled by changing the correlated electron and phonon  scattering strengths.

Now we outline main steps of our considerations. We consider the model of spinless fermions. This is sufficient since in the spinful case, where charge (c) and spin ($s$) degrees of freedom  are separated, low-energy  phonons are coupled to charge only. Thus, in the scaling dimension of the impurity term, $\Delta=\Delta_\mathrm{c}+\Delta_\mathrm{s}$, \cite{KF:92b} only   $\Delta_\mathrm{c}$ is affected by the $e$-ph coupling, while $\Delta_\mathrm{s}$ remains the same as in the phononless case so that calculating $\Delta_\mathrm{c}$ in the spinful case is effectively reduced to calculating $\Delta$ in the spinless case as follows.

The low-energy properties of spinless fermions can be described in terms  of bosonic fields $\theta_\mathrm{R,L}$  which parameterize density fluctuations of the right- and left-moving electrons, $2\pi \rho_\mathrm{R,L}({x})=\pm\partial_x\theta_\mathrm{R,L}({x})$. The spatial derivatives of their linear combinations,   canonically conjugate bosonic fields $\phi =\tfrac{1}{2}({\theta _\mathrm{R}+\theta _\mathrm{L}  })$ and  $\theta  = \tfrac{1}{2}({\theta _\mathrm{R}-\theta _\mathrm{L}  })$,  are proportional to the current and  the fluctuations of the full electron density, respectively. There is a duality between these fields: if $\theta$ is chosen as a generalized coordinate, then $\partial _x\phi $ plays the role of a generalized momentum, and vice versa \cite{Giamarchi,GogNersTsv}. Since both the   impurity and the  phonons are coupled only to the density, it is convenient to write the action of the TL model in the $\theta $-representation. Apart from the Luttinger parameter $K$, the model is characterized by the effective excitation (plasmon) velocity $v$ and the appropriate Lagrangian density in the Keldysh formalism \cite{Keldysh,*RS:86,*LevchKam} can be written as
\begin{align}
    \label{L0}
\mathcal{L} _0 &=\frac{1}{2\pi vK} \left\{{\left[\partial  _t\theta({\xi }) \right]^2-\left[v\partial  _x \theta(\xi )\right]^2}\right\} \,, &\xi&\equiv(x,t)\,.
\end{align}
In the dual representation,   the Lagrangian density has the same form as above  but with $\theta\to \varphi $ and $K\to1/K$.

Assuming the standard Debye model for one-dimensional acoustical phonons linearly coupled (with a coupling constant $g_\mathrm{ph} $) to the electron density adds, after integrating out phonon fields, the following (nonlocal and retarded) term to the Lagrangian density:
 \begin{align}
 \label{ph}
    &\mathcal{L}_{\mathrm{ph} }   =-\frac{\alpha v}{2\pi K }\partial _x\theta(\xi )\,  \mathcal{D}({t-t ' ; \,x,x'})\,\partial _{x'}\theta (\xi  ')\,,
\end{align}
where   $\alpha=g_\mathrm{ph}^2 K/\pi v$ is a dimensionless $e$-ph coupling constant and $\mathcal{D} $ is the phonon propagator. For a translationally invariant system (or when the impurity does not scatter phonons), $\mathcal{D}$ depends only on $x-x'$  and the retarded component of its Fourier transform  is given by the standard expression
\begin{align}\label{D0}
\mathcal{D}_0 ^{\mathrm{r}} (\omega, q)=\frac{c^2q^2}{(\omega+ i 0)^2-c^2q^2 }\,.
\end{align}
Here we do not consider a direct electron backscattering from phonons \cite{VoitSchulz,AndreevMatveev:05} since at low energies corrections to the effective $e$-$e$ interaction due to phonons with momentum $2k_F$ are local and nonretarding  \cite{VoitSchulz} and can be thus absorbed into redefined interaction constants. Instead, we focus on the electron coupling to acoustic phonons with low momenta  and its effect on the renormalization of the electron backscattering from an impurity. The latter is described by adding to the Lagrangian the usual term $\lambda \cos 2\theta (t)$, where $\lambda $ is a backscattering amplitude and $\theta({t})\equiv\theta({x=0,t})$.

Without the impurity, Eqs.~(\ref{L0}) and (\ref{ph}) (with $\mathcal{D}\to\mathcal{D}_0$) describe a two-component LL with excitation velocities $v_\pm$ given by $v_\pm^2/c^2 =\frac{1  }{2} [1+\beta^2 \pm \sqrt{({1-\beta^2})^2+4\alpha \beta^2}]$ \cite{Loss94,martin05}, where $\beta\equiv v/c$. We assume that $\alpha<1$ to avoid the Wentzel--Bardeen lattice instability \cite{Wentzel,*Bardeen:51} corresponding here to $v_-^2\le0$. Note in passing that a similar two-component propagation characterizes a fermion-boson mixture of cold atoms \cite{HoCaz,*F-BmixLutt}; embedding an impurity in such a mixture will be considered elsewhere.

If the  impurity breaks translational invariance for the phonon propagation, Eq.~(\ref{D0}) is not necessarily valid. However, it  remains applicable  in a relevant low-frequency limit when a lattice defect oscillates together with the 1D wire. In this case the phonon backscattering amplitude goes to zero at $\omega \to0$, whether the impurity effect on phonons is modeled by its mass or its spring constant being different from those on the lattice  \cite{martin05}.

On the contrary, phonons at $\omega \to0$ are fully reflected from the impurity  pinned to a substrate. In such a case they do not mediate between the electrons on different sides of the impurity, while the electrons on the same side feel both the direct and reflected phonons. Then a spatial structure of the phonon propagator in Eq.~(\ref{ph}) is $\mathcal{D}({x,x'}) = \left [{\mathcal{D}_0({x-x'})   + \mathcal{D}_0({x+x'}) }\right ] \Theta ({xx '}) $ (where $\Theta ({x})$ is the step function). Generalizing this for an arbitrary phonon scattering from the impurity, we write the retarded component of the phonon propagator as
\begin{align}\label{r}
 {\mathcal{D}}^{\mathrm{r}}({\omega; x,x'})= {\mathcal{D}}_0^{\mathrm{r}}({\omega; x\!-\!x'})
-r\sgn (xx') \,{\mathcal{D}}_0^{\mathrm{r}}({\omega;|x|\!+\!|x'|})\,,
\end{align}
implying that the scattering is described by a $2\times 2$ unitary matrix fully characterized by a (complex) reflection amplitude $r$ (with $r=-1$ corresponding to the full reflection limit above). Note that at $\omega=0$ translational invariance is either completely restored for the lattice defect (full transparency, i.e.\ $r({\omega\!=\!0 }) =0$) or broken for the pinned impurity  (full reflection, i.e.\ $r({\omega\!=\!0})=-1  $). However, phonon transmission at a relevant low-energy cutoff (e.g., $\omega _0\sim\max\{{T,\,c/L}\}  $ with $L$ being the wire length)  can, in principle, take an intermediate value. In the present Letter, we restrict ourselves to the case when $r$ is a real number between $0$ and $-1$.

The action corresponding to Eqs.~(\ref{L0})-({\ref{ph}}) is quadratic in the  fields $\theta ({x\ne0,t}) $. Integrating them out results in a nonlocal in time Lagrangian in terms of $\theta ({t})$:
\begin{align}\label{Lt}
    \mathcal{L}   &=\frac{1 }{2}\int\!\theta ({t})\,\mathcal{G} ^{-1}(t-t' )\,\theta ({t '}) \,\mathrm{d}t'- \lambda \cos( 2\theta (t) )\,.
\end{align}
Here  $\mathcal{G}({t-t'})\equiv\mathcal{G}({t-t'};\,x\!=\!0, x'\!=\!0)  $ is an autocorrelation function of the field $\theta ({t})$ in the presence of the $e$-ph coupling. A full Green function $\mathcal{G}({t-t'};\,x, x' ) $ describes collective excitations (polarons) in the two-component LL. It is convenient to parameterize the Fourier transform of the retarded component of $\mathcal{G}({t-t'})$ as
\begin{align}\label{P}
    \mathcal{G} ({\omega })&=-\frac{\pi i} {2}\frac1{\omega +i0}   \Delta ({\alpha ,\beta ,r })\,.
\end{align}
Without the $e$-ph coupling  $\Delta =K$ and Eqs.~(\ref{Lt})--({\ref{P}}) correspond to the effective $x=0$ action for the TL model with the impurity  \cite{KaneFis:92a} (but written here in the Keldysh formalism). A calculation of $ \Delta  ({\alpha ,\beta \, ,r })$ is outlined below.  Here we stress that it is just a number, which does not depend on $\omega $. Whatever is its value, the RG considerations of Ref.~\cite{KaneFis:92a} for the \emph{weak-scattering limit} remain valid so that calculating $\Delta ({\alpha ,\beta ,r })$ from Eqs.~(\ref{L0}) -- ({\ref{r}}) gives the scaling dimension $\Delta _\mathrm{WS} $ of $\lambda$ in this limit. Naturally, the presence of a local impurity does not renormalize values of $\alpha $ and $\beta $ in the bulk as well as it does not renormalize the value of $K$ \cite{Giamarchi}.

The $\lambda$-term in ({\ref{Lt}})  describes, in principle, backscattering of an arbitrary strength. Although the strong scattering limit can be treated using an instanton approximation \cite{FurusakiNagaosa:93b,Schmid:84}, an RG analysis of strong scattering can be done \cite{KaneFis:92a} by substituting the scattering  term  by a weak link  between two semi-infinite wires. This adds the tunneling term $t_\mathrm{WL} \cos 2\varphi  ({0},t)$ to the Lagrangian, with $\varphi  \equiv  [\phi_{l} -\phi  _ {r} ]  $  with the indices $l,r$ referring to the left and right sides of the wire. Without phonons, representing the action of the TL model in terms of   $\phi  $ instead of $\theta $ \cite{KaneFis:92a}, with the help of the duality between these fields described after Eq.~(\ref{L0}),  immediately results in the weak-link dimension $\Delta _\mathrm{WL}=1/K  $ and thus in the duality relation ({\ref{Dual}}).

In our case, when  the electron density fields $\partial _x\theta $ are coupled by the nonlocal phonon propagator (\ref{r}), expressing $\mathcal{L}_0$ in terms of the fields $\phi  $ would give no advantage  while require extra boundary conditions at $x=0$. Instead we use an ``unfolding'' procedure \cite{EggertAffleck:92,*EggertAffleck:95,*FabrizioGogolin:95} where nonchiral modes in each semi-infinite wire are mapped onto a  chiral mode in an  infinite wire. Then the weak tunneling between the two semi-infinite wires is mapped onto a weak scattering between the new chiral modes in the infinite wire. The inevitable loss of translational invariance in the interaction term resulting from  the unfolding is easy to cure \cite{Giamarchi} (in the absence of the $e$-ph coupling) by making the rescaling $\theta\to \theta\sqrt{K}$ (and $\phi \to \phi /\sqrt{K}$ to keep it canonically conjugate to $\theta$) before the unfolding. This removes
 the interaction by making $K$ equal to $1$. As a result, after the unfolding $\mathcal{L}_0$ retains  form (\ref{L0}) (but with $K=1$) in terms of the fields  $\widetilde\theta   $ (and $\widetilde\phi $) defined as the half-difference (and half-sum) of the  chiral fields resulted from the unfolding. The tunneling term after the rescaling and unfolding becomes $t_\mathrm{WL} \cos[2\widetilde\theta({t})/\sqrt{K}]$.

Although no rescaling can remove the phonon-mediated part of the action, Eq.~(\ref{ph}),   the action for an arbitrary phonon scattering from the impurity, Eq.~(\ref{r}), is not translationally invariant anyway.   Still the rescaling and unfolding procedure remains useful, albeit the resulting action becomes  rather complicated: the full electron density is not expressible via $\widetilde\theta$ alone and the phonon propagators thus couple the pairs of $\widetilde\theta$ and of $\widetilde\phi $. We perform the unfolding \cite{YGYL:1} using the mixed $\theta$-$\phi $ representation and integrate the fields $\widetilde\phi $ out afterwards.  After rescaling  $\widetilde\theta$  again, so that the tunnelling term becomes simply $t_\mathrm{WL} \cos[2\widetilde\theta({t}) ]$, the quadratic part of the Lagrangian density becomes
\begin{align}\label{swl}
\!\!\!\mathcal{L}_{\mathrm{WL}} =\frac{K}{2\pi v }\left[\partial_t\widetilde\theta({\xi }){  {\mathcal{Q}}}^{-1}\partial_{t'}\widetilde\theta({\xi '})-v ^2\partial_x\widetilde\theta({\xi })\widetilde {\mathcal{D}}\partial_{x'}\widetilde\theta ({\xi '})\right]
\end{align}
where the Fourier transforms of the retarded parts of the kernels $\widetilde {\mathcal{D}}$ and $\mathcal{Q}$ are expressed via $\mathcal{D}_0 $  in the mixed $\omega$-$x$ representation   as follows:
\begin{align} \notag
    \widetilde {\mathcal{D}}^{\mathrm{r}}&=\delta({x-x'})+\tfrac{1}{2}\alpha\left[  {\mathcal{D}}_0^{\mathrm{r}}({\omega; x-x'})+\mathcal{D}_0^{\mathrm{r}}({\omega;x+x'})\right],\\
    \mathcal{Q}^{\mathrm{r}}&=\widetilde {\mathcal{D}}^{\mathrm{r}}- \alpha({1+r})\mathcal{D}_0^{\mathrm{r}}({\omega; |x|+|x ' |}).\label{DQ}
\end{align}
As before, integrating out the fields $\widetilde \theta({x\ne0,t})$ results in the Lagrangian of the same form as in Eq.~(\ref{Lt}).

Calculating the Green function in Eq.~(\ref{Lt}) and thus the scaling dimensions in Eq.~(\ref{P}) requires  inverting the kernels of the Lagrangian densities of Eqs.\,(\ref{L0})-(\ref{ph}) for the WS case and of Eq.\,(\ref{swl}) for the WL case. Such an inversion, trivially done by a Fourier transform in a translationally invariant case, would not be possible for generic nonlocal kernels given by Eqs.~(\ref{r}) and (\ref{DQ}) due to the presence of $\mathcal{D}_0({\omega ;|x|+|x' |})$. The fact that it is possible in the present case is due to the factorizability, $\mathcal{D}_0^\mathrm{r} ({\omega; |x|+|x' |})=(2i/w_+)\mathcal{D}_0^\mathrm{r} ({\omega; |x|})\,\mathcal{D}_0^\mathrm{r} ({\omega; |x'|})$ (where $w_+\equiv\omega/c+i0$), which is ensured by the specific form of the propagator (\ref{D0}):  $\mathcal{D}_0^\mathrm{r} ({\omega; x})=- \frac{i}{2}w_+\mathrm{e}^{iw_+|x|}-\delta({x}) $. This makes the case of the additional interaction mediated by phonons (i.e.\ by excitations with a linear spectrum) rather special. Solving an integral equation for $\mathcal{G}$ in this case is straightforward, albeit cumbersome \cite{YGYL:1}, and leads to the nontrivial duality relation of Eq.~(\ref{Delta}). Note that this equation reproduces earlier results either for $\Delta_\mathrm{WS}$ at $r=0$ \cite{martin05,GYL:2011} or for $\Delta_{\mathrm{WL}} $ at $r=-1$ \cite{GYL:2011}.

It is worth stressing that building blocks for evaluating the  Green functions and thus $\Delta$ are rather different for the WS and WL cases. So it is quite surprising that the duality relation (\ref{Delta}) holds for an arbitrary set of parameters characterizing the $e$-ph coupling and phonon scattering from the impurity.

We reiterate the consequences of the duality: if electron backscattering from the impurity varies under the fixed values of all phonon parameters, the duality is directly applicable resulting in the phase diagram of Fig.~\ref{Fig1}. When the electron and phonon backscattering from the impurity are correlated, $\Delta _\mathrm{WS}({r\!=\!0}) $ and $\Delta _\mathrm{WL}({r\!=\!-1}) $ goes to $1$ at different values of $\alpha $ and $\beta $ resulting in the phase diagram of Fig.~\ref{Fig2}.

We do not have the evidence to decide whether the integrability of the standard TL model with an impurity \cite{FendleyLudwigSaleur:95} survives including an additional retarded interaction mediated by (not necessarily translationally invariant) bosons, or whether the duality exists for a broader range of nonintegrable 1D systems. Each of these possibilities is intriguing by itself. A possible way to rule the integrability out is to check whether the effective excitations decay in the presence of the impurity. Such a decay has been recently found in disordered Luttinger systems \cite{Bagrets:08} as well as in the pure Luttinger liquid in the presence of the spectral curvature \cite{Khodas:07}.  It seems plausible that adding phonons to the (otherwise integrable) TL models with a single impurity would allow a similar decay but to establish whether this is correct is a challenging task that certainly warrants further studies.

\begin{acknowledgments}
We gratefully acknowledge support from  the Leverhulme Trust via the Grant No.\ RPG-380 (I.V.Y.\ and I.V.L.) and the DFG through SFB TR-12 (O.M.Y.\ and I.V.Y.), as well as partial support from the Arnold Sommerfeld  Center for Nanoscience (I.V.Y.) and from the DOE Office of Science (A.G.) under the Contract No.\ DEAC02-06CH11357.
\end{acknowledgments}

 \newpage
 \addtolength{\textwidth}{.5in}
\addtolength{\textheight}{.5in}
\addtolength{\oddsidemargin}{-.9cm}
\setlength{\evensidemargin}{-.9cm}
 \begin{widetext}
 \renewcommand\theequation{S\arabic{equation}}
\def\sgn{\operatorname{sgn}}
\setcounter{equation}{0}
\centerline{\large \bf Supplemental Material}
\vspace*{12pt}

In the main text we discuss the robustness of the duality between weak and strong scattering when coupling to massless bosons (e.g., 1D phonons) is added to the standard Tomonaga-Luttinger model. The calculation of the scaling dimensions requires to find Green functions (resolvents) of the Lagrangian density given by Eqs.~(3) and (4) for the weak scattering case and by Eq.~(8) for the (dual) weak link case. Since the Lagrangian is quadratic, an appropriate procedure is rather straightforward and there is no need to describe it in the main text. On the other hand, it is not entirely standard, since the Lagrangian is not translationally invariant, and thus is worth describing here.

The critical exponent in both WS and WL limit is calculated from Eq.~(8) in the main text, which we rewrite below:
\begin{align}\label{G}
    \mathcal{G} ({\omega })&= \frac{\pi } {2i}\frac1{\omega +i0}   \Delta ({\alpha ,\beta \, ,r })\,.
\end{align}
Here $\mathcal{G} ({\omega })$ is the Fourier transform of the retarded component of the local Green function, $\mathcal{G}({t-t'};\,x\!=\!0, x'\!=\!0 ) $, which is a double integral over the momenta $q,\,q'$  of $\mathcal{G}(\omega ;\,q,q' ) $, the  Fourier transform of $\mathcal{G}({t-t'};\,x, x' ) $  with respect to $x$ and $x'$.

{\bf The weak scatterer}. The quadratic Lagrangian density is given by  $\mathcal{L}= \mathcal{L}_0+\mathcal{L}_\mathrm{ph}   $, Eqs.~(3) and (4) in the main text:
\begin{align}
    \label{L0}
 \mathcal{L}  =\frac{1}{2\pi vK} \left\{{\left[\partial  _t\theta({\xi }) \right]^2-\left[v\partial  _x \theta(\xi )\right]^2}
 - {\alpha v^2}\partial _x\theta(\xi )\,  \mathcal{D}({t-t ' ; \,x,x'})\,\partial _{x'}\theta (\xi  ')\right\} \,,\qquad \xi\equiv(x,t)\,.
\end{align}
We remind that $\alpha\equiv g_\mathrm{ph}^2 K/\pi v$ is the dimensionless constant  measuring relative strength of the el-ph coupling, $g_\mathrm{ph}$, and the el-el repulsion which increases when the Luttinger parameter $K$ decreases from $1$; we keep $\alpha<1$ to avoid the Wentzel--Bardeen instability. The Fourier transform yields
\begin{equation}\label{Gws}
    \mathcal{G}_{}^{-1} (\omega; q, q') = 2 \pi \delta(q - q') G_0^{-1} (\omega; q) -
          \frac{\alpha vqq'}{\pi K}   \mathcal{D} (\omega; q, q')  \,.
\end{equation}
Here $  G_0 (\omega; q) = \pi v K / \left[ \omega^2 - (v q)^2 \right]  $ is the plasmon Green function  in
the TL model, and $  \mathcal{D} (\omega; q, q')   $ is the Fourier transformed $   \mathcal{ D} (\omega; x, x') = {\mathcal{D}}_0 ({\omega; x\!-\!x'})
-r\sgn (xx') \,{\mathcal{D}}_0 ({\omega;|x|\!+\!|x'|})  $ (Eq.~6 in the main text), given by
\begin{equation}\label{D}
     \mathcal{D} (\omega; q, q') = 2 \pi \delta(q - q') \mathcal{D}_0 ( \omega; q)  - \frac{2 i \omega   r }{cqq'}
          {\mathcal{D}_0 ( \omega; q)} {\mathcal{D}_0 ( \omega; q')} \,,
\end{equation}
where all the propagators are retarded (i.e.\ $\omega\to\omega\!+\!i0$ where relevant) but the superscript in $ \mathcal{D}^{\mathrm r} $ is omitted. We reiterate that Eq.~(\ref{D}) has a simple form due to the factorizability, $\mathcal{D}_0  ({\omega; |x|+|x' |})\propto \mathcal{D}_0  ({\omega; |x|})\,\mathcal{D}_0  ({\omega; |x'|})$, where its Fourier transform is the standard phonon propagator, $\mathcal{D}_0({\omega ;\,q})=c^2q^2/(\omega ^2-c^2q^2) $.

It is straightforward to find $\mathcal{G}_\mathrm{}({\omega ;\,q, q'})  $ by inverting the r.h.s.\ of Eq.~(\ref{Gws}). For the sake of the weak-link case
below, we note that a more general kernel, $K(q,q') = 2\pi\delta(q-q') \, a_q + 2\pi\delta(q+q') \, b_q +c_qc_{q'},\,$ containing also a reflected part  proportional to $\delta ({q+q'})$, can be inverted as follows, provided that $a_q$ and $b_q$ are even functions of $q$:
\begin{align}
\label{InvKern}
    K^{-1}(q,q') &=
                2\pi\frac{a_q \delta(q-q')-b_q\delta(q+q')}{a^2_q-b^2_q}-
                \frac{1}{1+\int\frac{\mathrm{d}q}{2\pi}\frac{c^2_q}{a_q+b_q}}\,\frac{c_q}{a_q+b_q}\,\frac{c_{q'}}{a_{q'}+b_{q'}} \, ,
\end{align}
 As  $b_q=0$ for the weak-scattering propagator (\ref{Gws}) one easily finds its inverse
(where $\beta \equiv v/c$ is the ratio of the plasmon
 and phonon speeds):
\begin{align}\label{GwsQ}
    \frac{\mathcal{G}_{\mathrm{}}(\omega; q,q')}{\pi vK}&=G(\omega; q)\,2\pi\delta(q-q')
+
             \frac{2i\omega \, r \alpha \beta v }{1-2i\omega \, r \alpha \beta v \, \overline{ G\,\mathcal{D}^2 } } \,
             \mathcal{D}_0 ( \omega; q) G(\omega; q) \, \mathcal{D}_0 ( \omega; q') G(\omega; q') \,,\\\notag
             G(\omega; q) &\equiv \frac{1}{\omega^2-\left[1+\alpha \mathcal{D}_0 ( \omega; q) \right]v^2q^2}\,,\qquad
   \overline{ G\, \mathcal{D}^k} \equiv \int\frac{\mathrm{d}q}{2\pi}\, G(\omega; q)  \Bigl[ \mathcal{D}_0 ( \omega; q) \Bigr]^k\, .
\end{align}
In Eq.~(\ref{GwsQ}), $G(\omega; q) $ is the RPA polaron propagator for the LL liquid coupled to translationally invariant phonons, which goes over to the  plasmon propagator $G_0(\omega; q) $  for $\alpha=0$, while $\mathcal{G}(\omega; q,q')$ is the full RPA polaron propagator with allowance for the phonon reflection from impurity, Eq.~(\ref{D}). We emphasize that the RPA remains exact in the presence of the el-ph coupling.
The explicit expressions for $  \overline{ G \, \mathcal{D}^k}$, which we need at $k = 0, 1, 2  $, are found by calculating these pole integrals as follows:
\begin{align}\label{PoleInt}
   \overline{   G ^{\phantom 2\!\!} } &=
       \frac{1}{2i\omega}\,\frac{1\!+\!\beta \kappa}{v\,W \kappa}, &
    \overline{   G\mathcal{D} ^{\phantom 2\!\!}} &=
       -\frac{1}{2i\omega}\,\frac{1}{v\,W \kappa}, &
   \overline{ G \mathcal{D}^2} &=
     \frac{1}{2i\omega}\,\frac{1}{\alpha \beta v}\left[\frac{\kappa\!+\!\beta}{W \kappa}-1\right] ;&
\kappa&\equiv\sqrt{1\!-\!\alpha}\,,\; W\equiv\sqrt{1\!+\!2\beta \kappa\! +\! \beta^2}\,.
\end{align}
 Integrating both sides of Eq.(\ref{GwsQ}) over $q$ and $q'$ yields
\begin{equation*}
\frac{\mathcal{G}_{\mathrm{}}(\omega)}{\pi vK}=  \overline{ G ^{\phantom 2\!\!} }
+
             \frac{2i\omega \, r \alpha \beta v }{1-2i\omega \, r \alpha \beta v \, \overline{G \mathcal{D}^2 } } \,
             \left( \overline{  G\mathcal{D} ^{\phantom 2\!\!} } \right)^2 \, .
\end{equation*}
Substituting here expressions (\ref{PoleInt}), one sees that the $\omega$-dependence of the r.h.s.\ is reduced to an overall factor of $1/(i\omega)$. Then, comparing the result with Eq.~(\ref{G})  and restoring the subscript, we find
\begin{align}\label{Dws}
    \Delta_\mathrm{ws}&=\frac{K}{W\kappa}\left\{{1+\beta\kappa +\frac{r\alpha\beta}{(1+{r})W\kappa-{r} (\kappa+\beta)}}\right\} =K\,\frac{(1+{r})(1+\beta \kappa)-{r} W}{(1+{r})W\kappa-{r} (\kappa+\beta)}\,.
\end{align}
The last expression, which is easy to verify using the definitions of $\kappa$ and $W$, Eq.~(\ref{PoleInt}),  is given in the main text, Eq.~(2).

\vspace{0.5 cm}

{\bf The  weak link}. First we detail how to get the Lagrangian density of Eq.~(9) in the main text. The procedure was outlined there, but for completeness we repeat the steps described in the main text. Firstly, rescaling the fields as
$ \theta \to  \theta \, \sqrt{K} \,   $ and $
  \phi   \to   \phi   /  \sqrt{K} \,
$
results in the Lagrangian of the noninteracting TL model ({$K\!=\!1$}) with the tunneling term $t_\mathrm{wl} \cos [2\varphi  ({0},t)/\sqrt{K}] $ where $\varphi  \equiv  [\phi_{l} -\phi  _ {r} ]  $  with the indices $l,r$ referring to the left and right sides of the wire. We remind that $\phi  $ and $\theta $ are, respectively, half-sum and half-difference of the original chiral fields $\theta _\mathrm{L,R} $. The unfolding procedure introduces new chiral fields, $\widetilde\theta_\mathrm{R}(t,{x})=\theta_\mathrm{R}(t,{x})\Theta({x})+\theta_\mathrm{L}(t,{-x})\Theta({-x}) $ and
$\widetilde\theta_\mathrm{L}(t,{x})=\theta_\mathrm{R}(t,{-x})\Theta({x})+\theta_\mathrm{L}(t,{x})\Theta({-x}) $, where $\Theta({x})$ is the step function; these fields correspond to the old fields on the left and right sides from the impurity. Then we introduce fields $\widetilde{\theta }\equiv \frac{1}{2}({\widetilde{\theta }}_\mathrm{R}-\widetilde{\theta }_\mathrm{L}  ) $ and  $\widetilde{\phi   }\equiv \frac{1}{2}({\widetilde{\theta }}_\mathrm{R}+\widetilde{\theta }_\mathrm{L}  ) $. As a result, $\varphi  ({t})\to \widetilde{\theta} ({t})$ in the tunneling term, while the TL part of the Lagrangian retains the standard form in terms of $\widetilde{\theta }$ (the first two terms in Eq.~(\ref{L0}) but with $K=1$), since the interaction was effectively removed by the rescaling. However, since the el-ph part of the action depends after the unfolding on both $\widetilde{\theta }$ and $\widetilde{\phi  }$, it is convenient to  write both $\mathcal{L}_0 $ and $\mathcal{L}_\mathrm{ph}  $ in a mixed $\widetilde{\theta }$-$\widetilde{\phi  }$ representation as follows:
\begin{gather*}
    \mathcal{L}_0=-\frac{1}{\pi }\partial _t {\widetilde{\phi  }}   \partial _x {\widetilde{\theta }} -\frac{v}{2\pi } \left[ ({\partial _x\widetilde{\phi  }})^2\!+\!(\partial _x\widetilde{\theta })^2 \right]\,,\quad \mathcal{L}_\mathrm{ph}=
    -\frac{\alpha v}{2\pi }\left[ \partial _x\widetilde{\theta }({\xi })\,\mathcal{D}_+({\xi ,\xi '})\,\partial _{x'}\widetilde{\theta }   ({\xi '})+ \partial _x\widetilde{\phi   }({\xi })\,\widetilde{\mathcal{D}}_+({\xi ,\xi '})\,\partial _{x'}\widetilde{\phi   }   ({\xi '}) \right] ,\\
    \mathcal{D}_+({\xi ,\xi '})=\frac{1}{2} \left[ \mathcal{D}_0({\omega ; x-x'}) +\mathcal{D}_0({\omega ; x+x'})  \right]\,,\qquad    \widetilde{\mathcal{D}}_+({\xi ,\xi '})= \mathcal{D}_+({\xi ,\xi '}) -({1+r}) \mathcal{D}_0({\omega ;| x|+|x|'}) \,.
\end{gather*}
Since the full action corresponding to $\mathcal{L}_0+\mathcal{L}_\mathrm{ph}+\mathcal{L}_\mathrm{tun}     $ is quadratic in $\phi  $, integrating this field out and rescaling again $\widetilde\theta\to \widetilde{\theta }\sqrt{K} $ results in the action with  the quadratic part of the Lagrangian density given by Eq.~({8}) in the main text. We rewrite its kernel as
\begin{equation}\label{Gwl}
   \mathcal{G}_{\mathrm{wl}}^{-1} (\omega; q, q') = \frac{K}{\pi v}
     \left[
       \omega^2 Q^{-1}(\omega; q, q') - v^2 q \widetilde{\mathcal{D}}(\omega; q, q') q'
     \right] \, ,
\end{equation}
where the Fourier transform of the propagators $\widetilde {\mathcal{D}}$ and $\mathcal{Q}$, Eq.~(9) in the main text, are given by
\begin{align} \notag
    \widetilde {\mathcal{D}}^{} (\omega; q, q') &= 2 \pi \delta(q - q') +
               \pi \alpha \Bigl[
            {\mathcal{D}}_0^{}({\omega; q}) \delta(q - q')+\mathcal{D}_0^{}({\omega;q}) \delta(q+q')
                               \Bigr] \, , \\
    \mathcal{Q}^{}(\omega; q, q') &=\widetilde {\mathcal{D}}^{}(\omega; q, q') -
               2 i \, \alpha({1+r}) \left( \frac{\omega}{c} \right)^3
                 \frac{\mathcal{D}_0^{}({\omega; q})}{(q)^2}
                 \frac{\mathcal{D}_0^{}({\omega; q'})}{(q')^2} \, .
\label{DQ-qRepr}
\end{align}
The kernel $\mathcal{G}^{-1} $ in Eqs.~(\ref{Gwl})--(\ref{DQ-qRepr}) has the structure that can be inverted with the help of Eq.~(\ref{InvKern}):
\begin{eqnarray}\notag
   \frac{ \mathcal{G}_{\mathrm{wl}}(\omega; q,q') }{\pi v K^{-1}} & = &
     G(\omega; q) \Bigl[1+\alpha \mathcal{D}_0 (\omega; q) \Bigr] \,2\pi\delta(q-q') - \\
     & - &
     \frac{ 2i\omega \, (1+r) \alpha c }{1+ 2i\omega \, (1+r)  \alpha \beta v\left( \overline{ G\mathcal{D}^{\phantom1\!\!}} + \overline{ G \mathcal{D}^2} \right) }
     \Bigl[ 1+ \mathcal{D}_0 (\omega; q) \Bigr] G(\omega;q) \, \Bigl[ 1+ \mathcal{D}_0 (\omega; q') \Bigr] G(\omega;q') \, .
\nonumber
\end{eqnarray}
Integrating this over both momenta, we find the local Green function
\begin{equation}\label{GwlLocal}
   \frac{ \mathcal{G}_{\mathrm{wl}}(\omega) }{\pi vK^{-1}} = \overline{  G^{\phantom1\!\!}} + \alpha   \overline{ G\mathcal{D}^{\phantom1\!\!}}  -
     \frac{ 2i\omega  (1+r)  \alpha c }{1+ 2i\omega \, (1+r)  \alpha \beta v\left(\overline{ G\mathcal{D}^{\phantom1\!\!}}  + \overline{ G \mathcal{D}^2} \right) }
     \left( \overline{ G^{\phantom1\!\!}}  + \overline{ G\mathcal{D}^{\phantom1\!\!}}  \right)^2 \, .
\end{equation}
Substituting here the pole integrals (\ref{PoleInt}) and using the relation (\ref{G}) for the scaling dimension, we find
\begin{align}\label{Dwl}
    \Delta_\mathrm{wl}&=\frac1{KW}\left[ {\kappa+\beta } - \frac{({1+r})\alpha\beta}{W+({1+r})({1+\beta\kappa -W})}\right]\,.
\end{align}
Using the definitions of $W$ and $\kappa$, Eq.~(\ref{PoleInt}), it is straightforward to verify that the product of the l.h.s.\ of Eqs.~(\ref{Dwl}) and (\ref{Dws}) is, indeed, identically equal to 1.

\end{widetext}
\end{document}